%% file: nlo3jet-prl.tex
\documentclass[reprint, superscriptaddress, showpacs,preprintnumbers,
  nobibnotes, amsmath,amssymb, aps]{revtex4-1}

\newcommand{\GeV}{\ensuremath{\,\mathrm{GeV}}\xspace}
\newcommand{\TeV}{\ensuremath{\,\mathrm{TeV}}\xspace}

\usepackage{xspace}
\usepackage{graphicx}
\usepackage{color}
\usepackage{dcolumn}
\usepackage{bm}

\begin{document}

\preprint{DESY 13-144, FTUV-13-1308, IFIC/13-55, LPN13-055, LU-TP 13-28, MAN/HEP/2013/17}

\title{Electroweak Higgs plus Three Jet Production at NLO QCD}

\author{Francisco Campanario}
\affiliation{Theory Division,
  IFIC, University of Valencia-CSIC, E-46100 Paterna, Valencia,
  Spain}
\author{Terrance M. Figy}
\affiliation{School of Physics and Astronomy, University of Manchester}
\author{Simon Pl\"atzer}
\affiliation{Theory Group, DESY Hamburg}
\author{Malin Sj\"odahl}
\affiliation{Dept. of Astronomy and Theoretical Physics, Lund University, 
  S\"olvegatan 14A, 223\,62~Lund, Sweden}
\begin{abstract}
We calculate next-to-leading order (NLO) QCD corrections to
electroweak Higgs plus three jet production. Both vector boson fusion
(VBF) and Higgs-strahlung type contributions are included along with
all interferences. The calculation is implemented within the Matchbox
NLO framework of the Herwig++ event generator.
\end{abstract}


\maketitle

\section{Introduction}
Higgs production via vector boson fusion~(VBF) is an essential channel
at the LHC for disentangling the Higgs boson's coupling to fermions and
gauge bosons.  The observation of two tagging jets is crucial to
reduce the backgrounds. Furthermore, the possibility to identify Higgs
production via VBF is enhanced by being able to increase the signal to
background ratio by vetoing additional soft radiation in the central
region~\cite{Barger:1994zq,Rainwater:1998kj,Rainwater:1999sd,DelDuca:2004wt,Forshaw:2007vb,Andersen:2008gc,Cox:2010ug},
since this suppresses important QCD backgrounds, including $Hjj$ via
gluon fusion.

To exploit the central jet veto~(CJV) strategy for Higgs coupling
measurements, the reduction factor caused by the CJV on the observable
signal must be accurately known. The fraction of VBF Higgs events with
at least one additional veto jet between the tagging jets provides the
relevant information, i.e., we need to know the ratio of $ H j j j$
production to the inclusive cross section of $H j j$ production via
VBF. Gluon fusion NLO QCD corrections for $ H j j j$ within the top
effective theory approximation have been recently computed
in~\cite{Cullen:2013saa}, a validation of the effective theory
approximation at LO for this process can be found
in~\cite{Campanario:2013mga}.

\begin{figure}[ht]
\includegraphics[scale=0.7]{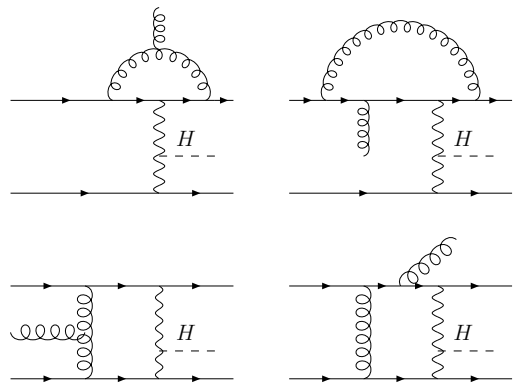}
\caption{\label{fig:ggf}
 Representative examples of diagrams of $Hjjj$ production via vector
 boson fusion.}  
\end{figure}

At present, the NLO QCD corrections of $ H j j j$ via VBF were
calculated~\cite{Figy:2007kv} with several approximations and without
the inclusion of five- and six-point function diagrams
(Fig.~\ref{fig:ggf}, second row) and the corresponding real emission
cuts, which were estimated to contribute at the per mille level.
However, some studies~\cite{Forshaw:2007vb} suggest that the
contribution of the missing pieces can be larger.  Due to the
importance of the process to Higgs measurements, a full calculation is
necessary to ensure that integrated cross sections and kinematic distributions
are not underestimated.

In this letter, we present results for electroweak Higgs plus three
jet production at NLO QCD. In Section~\ref{sec:caldetails}, we present
technical details of our computation.  In Section~\ref{sec:numres}, we
present numerical results and the impact of the NLO QCD corrections on
various differential distributions. Finally, we summarize our findings
in Section~\ref{sec:concl}.

\section{Calculational Details}
\label{sec:caldetails}

For the calculation of the LO matrix elements for Higgs plus two (we
have performed the $Hjj$ calculation in parallel within the same
framework), three and four jets, we have used the builtin spinor
helicity library of the Matchbox module \cite{Platzer:2011bc} to build
up the full amplitude from hadronic currents.  For the Born-virtual
interference the helicity amplitude methods described
in~\cite{Hagiwara:1988pp} have been employed. Both methods resulted in
two different implementations of the tree level amplitudes which have
been validated against each other, including color correlated matrix
elements.  The LO implementation has further been cross checked
against Sherpa~\cite{Gleisberg:2003xi,Gleisberg:2008ta} and
Hawk~\cite{Ciccolini:2007jr,Ciccolini:2007ec}.  The dipole subtraction
terms \cite{Catani:1996vz} have been generated automatically by the
Matchbox module~\cite{Platzer:2011bc}, which is also used for a
diagram-based multichannel phase space generation. For the virtual
corrections, we have used in-house routines, extending the techniques
developed in~\cite{Campanario:2011cs}. The amplitudes have been cross
checked against GoSam~\cite{Cullen:2011ac}.  A representative set of
topologies contributing are depicted in Fig.~\ref{fig:ggf}.

For electroweak propagators, we have introduced finite width effects
following \cite{Denner:2006ic}, using complex gauge boson masses and a
derived complex value of the sine of the weak mixing angle. We used
the OneLoop library~\cite{vanHameren:2010cp} which supports complex
masses to calculate the scalar integrals. For the reduction of the
tensor coefficients up to four-point functions, we apply the
Passarino-Veltman approach \cite{Passarino:1978jh}, and for the
numerical evaluation of the five and six point coefficients, we use
the Denner-Dittmaier scheme~\cite{Denner:2005nn}, following the layout
and notation of~\cite{Campanario:2011cs}.

To ensure the numerical stability of our code, we have implemented
a test based on Ward identities~\cite{Campanario:2011cs}. These Ward
identities are applied to each phase space point and diagram, 
at the expense of a small additional computing time and using a cache system. 
If the identities are not fulfilled, the amplitudes of gauge related
topologies are set to zero. The occurrence of these instabilities are
at the per-mille level, and therefore well under control. This method
was also successfully applied in other two to four
processes~\cite{Campanario:2011ud,Campanario:2013qba}. In the present
work, it is applied for the first time to a process which involves
loop propagators with complex masses.

The color algebra has been performed using ColorFull
\cite{Sjodahl:ColorFull} and double checked using ColorMath
\cite{Sjodahl:2012nk}. Within the same framework, we have implemented
the corresponding calculation of electroweak $Hjj$ production and
performed cross checks against
Hawk~\cite{Ciccolini:2007jr,Ciccolini:2007ec}.

\section{Numerical Results}
\label{sec:numres}

In our calculation, we choose $m_Z=91.188 \GeV$, $m_W=80.419002 \GeV$,
$m_H=125 \GeV$ and $G_F=1.16637\times 10^{-5}\GeV^{-2}$ as electroweak
input parameters and derive the weak mixing angle $\sin \theta_{W}$
and $\alpha_{QED}$ from standard model tree level relations. All
fermion masses (except the top quark) are set to zero and effects from
generation mixing are neglected.  The widths are calculated to be
$\Gamma_W=2.0476$ GeV and $\Gamma_Z=2.4414$ GeV.  We use the
CT10~\cite{Lai:2010vv} parton distribution functions with
$\alpha_s(M_Z)= 0.118$ at NLO, and the CTEQ6L1
set~\cite{Pumplin:2002vw} with $\alpha_s(M_Z)=0.130$ at LO. We use the
five-flavor scheme and the center-of-mass energy is fixed to
$\sqrt{s} = 14 \TeV$.

To study the impact of the QCD corrections, we use minimal inclusive
cuts.  We cluster jets with the anti-$k_T$
algorithm~\cite{Cacciari:2008gp} using {\tt FastJet}
\cite{Cacciari:2011ma} with $D=0.4$, $E$-scheme recombination and
require at least three jets with transverse momentum $p_{T,j} \ge
20~\rm{GeV}$ and rapidity $|y_{j}| \le 4.5$. Jets are ordered in
decreasing transverse momenta.

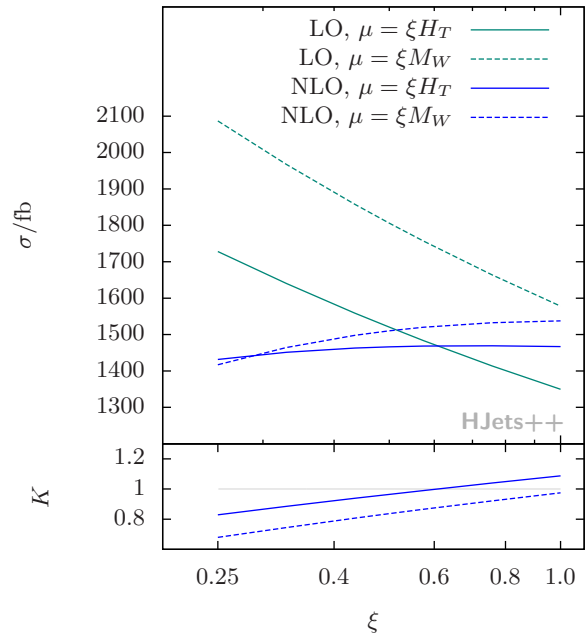
\begin{figure}[ht]
\begin{center}
\hspace*{2.3cm}\input{InclusiveMuVariation}
\end{center}
\caption{\label{fig:scale} The $Hjjj$ inclusive total cross section
  (in fb) at LO (cyan) and at NLO (blue) for the scale choices,
  $\mu=\xi M_{W}$ (dashed) and $\mu=\xi H_{T}$ (solid).  We also show
  the $K$-factor, $K=\sigma_{NLO}/\sigma_{LO}$ for $\mu=\xi M_{W}$
  (dashed) and $\mu=\xi H_{T}$ (solid). }
\end{figure}

In Figure~\ref{fig:scale}, we show the LO and NLO total cross-sections
for inclusive cuts for different values of the factorization and
renormalization scale varied around the central scale, $\mu$ for two
scale choices, $M_W/2$, and the scalar sum of the jet transverse
momenta, $\mu_{R}=\mu_{F}= \mu= H_{T}/2$ with $H_{T}=\sum_{j}
p_{T,j}$. In general, we see a somewhat increased cross section and -
as expected - decreased scale dependence in the NLO results.  We also
note that the central values for the various scale choices are closer
to each other at NLO.  The uncertainties obtained by varying the
central value a factor two up and down are around $30\%$ ($24\%$) at LO and
$2\%$ ($9\%$) at NLO using $H_{T}/2$ ($M_W/2$) as scale choice. 
At $\mu = H_T/2$, we obtained
$\sigma_{LO}=1520(8)^{+208}_{-171}$ fb
$\sigma_{NLO}=1466(17)^{+1}_{-35}$ fb.  Studying differential
distributions, we find that these generally vary less using the scalar
transverse momentum sum choice, used from now on.

\begin{figure}[ht]
\includegraphics[scale=0.7]{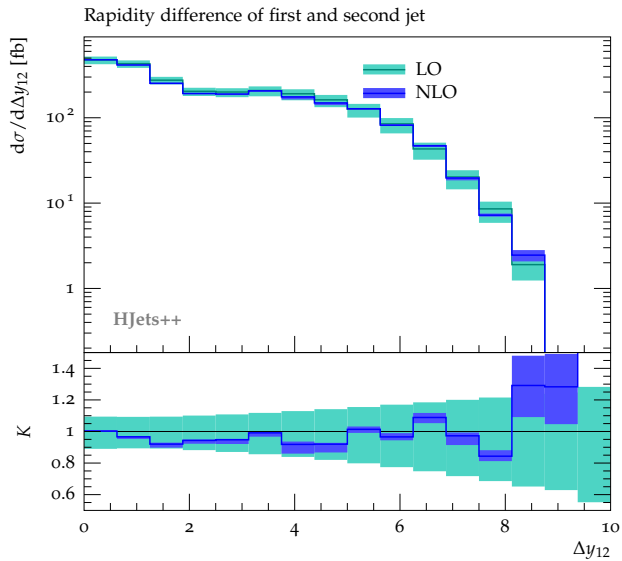}
\caption{\label{fig:j12dy} Rapidity difference differential
  distribution of the leading jets. Cuts are described in the
  text. The bands correspond to varying $\mu_F=\mu_R$ by factors 1/2
  and 2 around the central value $H_{T}/2$.  }
 \end{figure}

\begin{figure}[ht]
\includegraphics[scale=0.7]{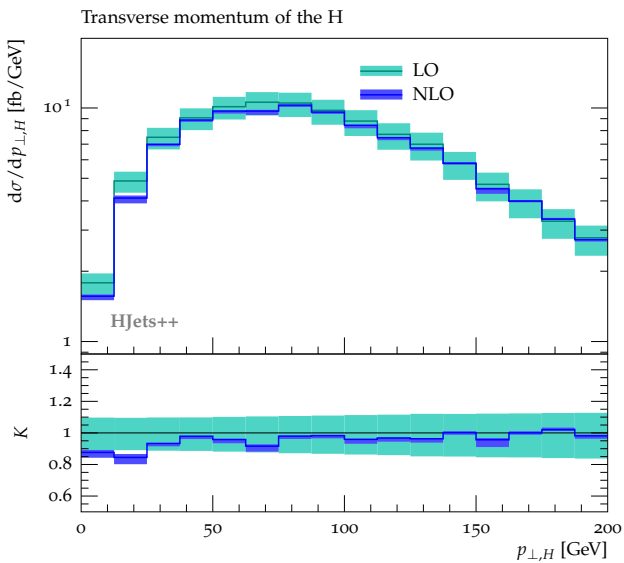}
\caption{\label{fig:hpt} Differential cross section and K factor for
  the $p_T$ of the Higgs. Cuts are described in the text. The bands
  correspond to varying $\mu_F=\mu_R$ by factors 1/2 and 2 around the
  central value $H_{T}/2$.  }
\end{figure}

In the following, we show some of the differential distributions which
characterize the typical vector boson fusion selection cuts and
central jet veto strategies as well as the $p_T$ spectrum of the
Higgs.

For the leading jets (defined to be the two jets with highest
transverse momenta $ p_{T,1}$ and $ p_{T,2}$), we show the rapidity
difference in Fig.~\ref{fig:j12dy}.  Generally, we find small
differences in shape compared to the LO results.

In Figure~\ref{fig:hpt}, the differential distribution for the $p_T$
of the Higgs is shown.  The NLO corrections are moderate over the
whole spectrum and the scale uncertaities are clearly smaller.

\begin{figure}[ht]
\includegraphics[scale=0.7]{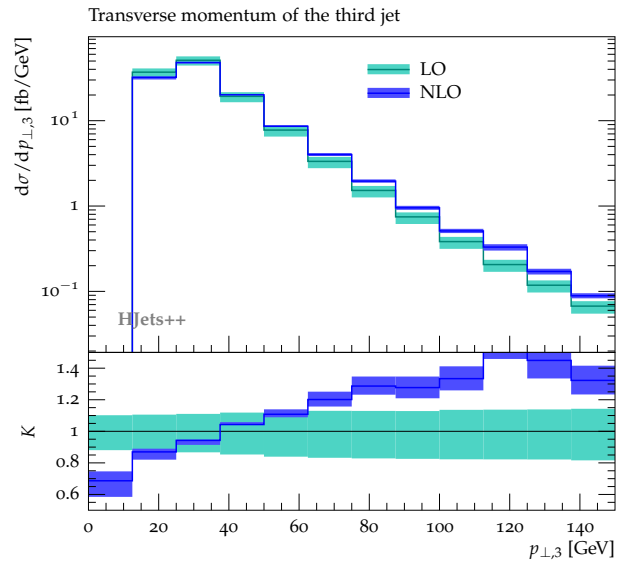}
\caption{\label{fig:pt3} Differential cross section and K factor for
  the $p_T$ of the third hardest jet. Cuts are described
  in the text. The bands correspond to varying $\mu_F=\mu_R$ by
  factors 1/2 and 2 around the central value $H_{T}/2$.  }
\end{figure}

In Figure~\ref{fig:pt3}, the differential distribution of the third
jet, the vetoed jet for a CJV analysis, is presented. Here we find
large differences in the high energy tail of the transverse momentum
distribution. Such high energy jets are significantly enhanced at NLO.

We also study the normalized centralized rapidity distribution of the
third jet w.r.t. the tagging jets,
$z^{*}_{3}=(y_3-\frac{1}{2}(y_1+y_2))/(y_1-y_2)$. This variable
beautifully displays the VBF nature present in the process.  One
clearly sees how the third jet tends to accompany one of the leading
jets appearing at $1/2$ and $-1/2$ respectively. This effect is
expected to be yet more pronounced when VBF cuts are applied, and
should be contrasted with the gluon fusion production mechanism where
radiation in the mid-rapidity will be much more common due to the
$t$-channel color flow of the process
\cite{Forshaw:2007vb,Cox:2010ug,Campanario:2013mga,Cullen:2013saa}.
Note also that the ordering of the leading jets in transverse momenta
generates an asymmetric radiation pattern. There is more radiation
around $z_3^*=1/2$ corresponding to emission of the third jet along
the first leading jet, $y_3=y_1$. Additionally, we remark, however,
that studying the rapidity distribution of the third jet alone does
not display the same characteristic pattern; here the peaks
accompanying the forward jets in the VBF region are completely smeared
out~(not shown here).

\begin{figure}[ht]
\includegraphics[scale=0.7]{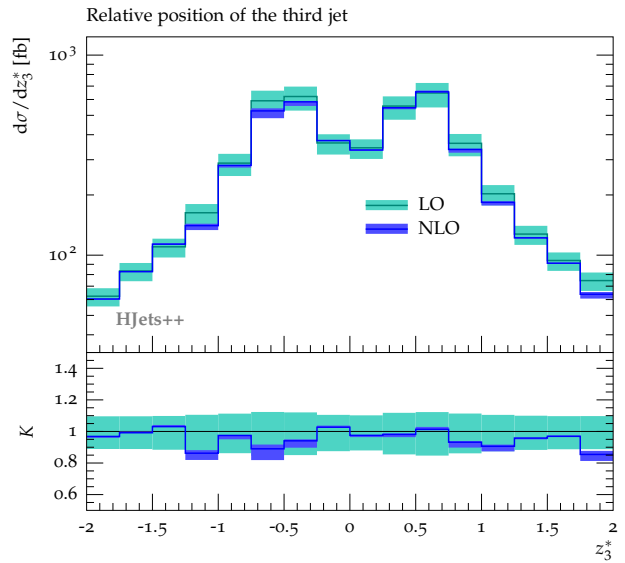}
\caption{\label{fig:z3} Differential cross section and K factor for
  the normalized centralized rapidity distribution of the third jet
  w.r.t. the tagging jets. Cuts are described in the text. The bands
  correspond to varying $\mu_F=\mu_R$ by factors 1/2 and 2 around the
  central value $H_{T}/2$.  }
\end{figure}

\section{Conclusions}
\label{sec:concl}
In this letter, complete results at NLO QCD order for electroweak
Higgs production in association with three jets have been
presented. The NLO corrections to the total inclusive cross section
are moderate for the inclusive cuts and a central scale of $H_T/2$
considered here, but can be more significant for a scale choice of
$M_W/2$. However, they exhibit a non-trivial phase space dependence
reflected in non-constant differential K-factors, particularly for jet
$p_T$ spectra. The scale uncertainty significantly decreases from
around $30\%$ ($24\%$) at LO down to about $2\%$ ($9\%$) at NLO 
using $H_T/2$ ($M_W/2$).

Despite our inclusive cuts and the possibility to radiate jets in the
central region, Higgs + 3 jets prefer VBF kinematics as seen in the
$z^{*}$ differential distribution. A detailed discussion of VBF and
jet vetoes will be given elsewhere.  Results including the effect of
parton showering are also anticipated.  The code is available upon
request and will be publicly available in the near future.

\section*{Acknowledgements}
We are grateful to Ken Arnold for contributions at an early stage of
this project and to Mike Seymour and Jeff Forshaw for valuable
discussions on the subject.  F.C. is funded by a Marie Curie
fellowship (PIEF-GA-2011-298960) and partially by MINECO
(FPA2011-23596) and by LHCPhenonet (PITN-GA-2010-264564).  T.F. would
like to thank the North American Foundation for The University of
Manchester and George Rigg for their financial support.  S.P. has been
supported in part by the Helmholtz Alliance ``Physics at the
Terascale'' and M.S. was supported by the Swedish Research Council,
contract number 621-2010-3326.

\bibliography{nlo3jet-prl}

\end{document}

%% file: InclusiveMuVariation.tex
\begingroup
  \makeatletter
  \providecommand\color[2][]{%
    \GenericError{(gnuplot) \space\space\space\@spaces}{%
      Package color not loaded in conjunction with
      terminal option `colourtext'%
    }{See the gnuplot documentation for explanation.%
    }{Either use 'blacktext' in gnuplot or load the package
      color.sty in LaTeX.}%
    \renewcommand\color[2][]{}%
  }%
  \providecommand\includegraphics[2][]{%
    \GenericError{(gnuplot) \space\space\space\@spaces}{%
      Package graphicx or graphics not loaded%
    }{See the gnuplot documentation for explanation.%
    }{The gnuplot epslatex terminal needs graphicx.sty or graphics.sty.}%
    \renewcommand\includegraphics[2][]{}%
  }%
  \providecommand\rotatebox[2]{#2}%
  \@ifundefined{ifGPcolor}{%
    \newif\ifGPcolor
    \GPcolortrue
  }{}%
  \@ifundefined{ifGPblacktext}{%
    \newif\ifGPblacktext
    \GPblacktexttrue
  }{}%
  \let\gplgaddtomacro\g@addto@macro
  \gdef\gplbacktext{}%
  \gdef\gplfronttext{}%
  \makeatother
  \ifGPblacktext
    \def\colorrgb#1{}%
    \def\colorgray#1{}%
  \else
    \ifGPcolor
      \def\colorrgb#1{\color[rgb]{#1}}%
      \def\colorgray#1{\color[gray]{#1}}%
      \expandafter\def\csname LTw\endcsname{\color{white}}%
      \expandafter\def\csname LTb\endcsname{\color{black}}%
      \expandafter\def\csname LTa\endcsname{\color{black}}%
      \expandafter\def\csname LT0\endcsname{\color[rgb]{1,0,0}}%
      \expandafter\def\csname LT1\endcsname{\color[rgb]{0,1,0}}%
      \expandafter\def\csname LT2\endcsname{\color[rgb]{0,0,1}}%
      \expandafter\def\csname LT3\endcsname{\color[rgb]{1,0,1}}%
      \expandafter\def\csname LT4\endcsname{\color[rgb]{0,1,1}}%
      \expandafter\def\csname LT5\endcsname{\color[rgb]{1,1,0}}%
      \expandafter\def\csname LT6\endcsname{\color[rgb]{0,0,0}}%
      \expandafter\def\csname LT7\endcsname{\color[rgb]{1,0.3,0}}%
      \expandafter\def\csname LT8\endcsname{\color[rgb]{0.5,0.5,0.5}}%
    \else
      \def\colorrgb#1{\color{black}}%
      \def\colorgray#1{\color[gray]{#1}}%
      \expandafter\def\csname LTw\endcsname{\color{white}}%
      \expandafter\def\csname LTb\endcsname{\color{black}}%
      \expandafter\def\csname LTa\endcsname{\color{black}}%
      \expandafter\def\csname LT0\endcsname{\color{black}}%
      \expandafter\def\csname LT1\endcsname{\color{black}}%
      \expandafter\def\csname LT2\endcsname{\color{black}}%
      \expandafter\def\csname LT3\endcsname{\color{black}}%
      \expandafter\def\csname LT4\endcsname{\color{black}}%
      \expandafter\def\csname LT5\endcsname{\color{black}}%
      \expandafter\def\csname LT6\endcsname{\color{black}}%
      \expandafter\def\csname LT7\endcsname{\color{black}}%
      \expandafter\def\csname LT8\endcsname{\color{black}}%
    \fi
  \fi
  \setlength{\unitlength}{0.0500bp}%
  \begin{picture}(4320.00,5544.00)%
    \gplgaddtomacro\gplbacktext{%
      \csname LTb\endcsname%
      \put(-106,1775){\makebox(0,0)[r]{\strut{} 1300}}%
      \put(-106,2049){\makebox(0,0)[r]{\strut{} 1400}}%
      \put(-106,2324){\makebox(0,0)[r]{\strut{} 1500}}%
      \put(-106,2598){\makebox(0,0)[r]{\strut{} 1600}}%
      \put(-106,2873){\makebox(0,0)[r]{\strut{} 1700}}%
      \put(-106,3148){\makebox(0,0)[r]{\strut{} 1800}}%
      \put(-106,3422){\makebox(0,0)[r]{\strut{} 1900}}%
      \put(-106,3697){\makebox(0,0)[r]{\strut{} 2000}}%
      \put(-106,3971){\makebox(0,0)[r]{\strut{} 2100}}%
      \put(3025,1280){\makebox(0,0){\strut{}}}%
      \put(-1008,3147){\rotatebox{-270}{\makebox(0,0){\strut{}$\sigma$/fb}}}%
    }%
    \gplgaddtomacro\gplfronttext{%
      \csname LTb\endcsname%
      \put(2216,4622){\makebox(0,0)[r]{\strut{}LO, $\mu=\xi H_{T}$}}%
      \csname LTb\endcsname%
      \put(2216,4402){\makebox(0,0)[r]{\strut{}LO, $\mu=\xi M_{W}$}}%
      \csname LTb\endcsname%
      \put(2216,4182){\makebox(0,0)[r]{\strut{}NLO, $\mu=\xi H_{T}$}}%
      \csname LTb\endcsname%
      \put(2216,3962){\makebox(0,0)[r]{\strut{}NLO, $\mu=\xi M_{W}$}}%
      \csname LTb\endcsname%
      \put(2663,1665){\makebox(0,0){\definecolor{Gray}{gray}{.7}\textcolor{Gray}{\sffamily\bfseries HJets++}}}%
    }%
    \gplgaddtomacro\gplbacktext{%
      \put(-106,931){\makebox(0,0)[r]{\strut{} 0.8}}%
      \put(-106,1159){\makebox(0,0)[r]{\strut{} 1}}%
      \put(-106,1386){\makebox(0,0)[r]{\strut{} 1.2}}%
      \put(442,484){\makebox(0,0){\strut{}0.25}}%
      \put(1318,484){\makebox(0,0){\strut{}0.4}}%
      \put(2073,484){\makebox(0,0){\strut{}0.6}}%
      \put(2610,484){\makebox(0,0){\strut{}0.8}}%
      \put(3025,484){\makebox(0,0){\strut{}1.0}}%
      \put(-876,1102){\rotatebox{-270}{\makebox(0,0){\strut{}$K$}}}%
      \put(1614,154){\makebox(0,0){\strut{}$\xi$}}%
    }%
    \gplgaddtomacro\gplfronttext{%
    }%
    \gplbacktext
    \put(0,0){\includegraphics{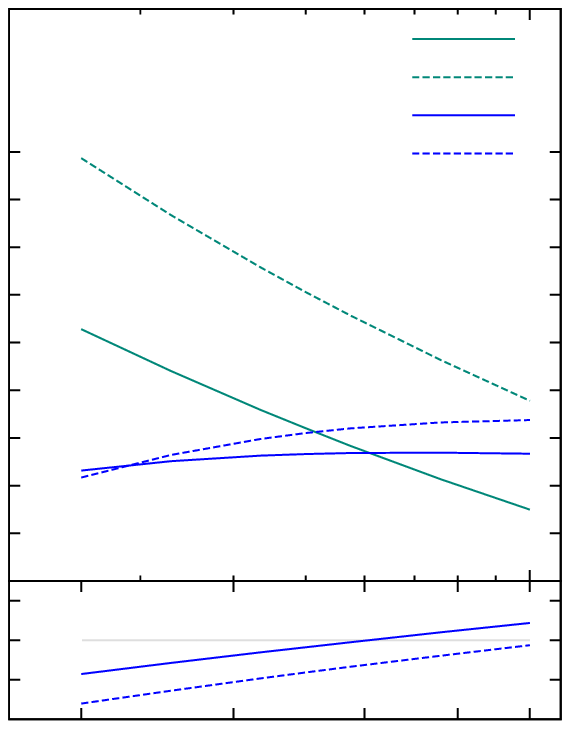}}%
    \gplfronttext
  \end{picture}%
\endgroup

%% file: nlo3jet-prl.bbl
\begin{thebibliography}{10}

\bibitem{Barger:1994zq}
V.~D. Barger, R.~Phillips, and D.~Zeppenfeld,
\newblock Phys.Lett. {\bf B346}, 106 (1995), hep-ph/9412276.

\bibitem{Rainwater:1998kj}
D.~L. Rainwater, D.~Zeppenfeld, and K.~Hagiwara,
\newblock Phys.Rev. {\bf D59}, 014037 (1998), hep-ph/9808468.

\bibitem{Rainwater:1999sd}
D.~L. Rainwater and D.~Zeppenfeld,
\newblock Phys.Rev. {\bf D60}, 113004 (1999), hep-ph/9906218.

\bibitem{DelDuca:2004wt}
V.~Del~Duca, A.~Frizzo, and F.~Maltoni,
\newblock JHEP {\bf 0405}, 064 (2004), hep-ph/0404013.

\bibitem{Forshaw:2007vb}
J.~R. Forshaw and M.~Sjodahl,
\newblock JHEP {\bf 0709}, 119 (2007), 0705.1504.

\bibitem{Andersen:2008gc}
J.~R. Andersen, V.~Del~Duca, and C.~D. White,
\newblock JHEP {\bf 0902}, 015 (2009), 0808.3696.

\bibitem{Cox:2010ug}
B.~E. Cox, J.~R. Forshaw, and A.~D. Pilkington,
\newblock Phys.Lett. {\bf B696}, 87 (2011), 1006.0986.

\bibitem{Cullen:2013saa}
G.~Cullen {\em et~al.},
\newblock (2013), 1307.4737.

\bibitem{Campanario:2013mga}
F.~Campanario and M.~Kubocz,
\newblock (2013), 1306.1830.

\bibitem{Figy:2007kv}
T.~Figy, V.~Hankele, and D.~Zeppenfeld,
\newblock JHEP {\bf 0802}, 076 (2008), 0710.5621.

\bibitem{Platzer:2011bc}
S.~Platzer and S.~Gieseke,
\newblock (2011), 1109.6256.

\bibitem{Hagiwara:1988pp}
K.~Hagiwara and D.~Zeppenfeld,
\newblock Nucl.Phys. {\bf B313}, 560 (1989).

\bibitem{Gleisberg:2003xi}
T.~Gleisberg {\em et~al.},
\newblock JHEP {\bf 02}, 056 (2004), hep-ph/0311263.

\bibitem{Gleisberg:2008ta}
T.~Gleisberg {\em et~al.},
\newblock JHEP {\bf 02}, 007 (2009), 0811.4622.

\bibitem{Ciccolini:2007jr}
M.~Ciccolini, A.~Denner, and S.~Dittmaier,
\newblock Phys.Rev.Lett. {\bf 99}, 161803 (2007), 0707.0381.

\bibitem{Ciccolini:2007ec}
M.~Ciccolini, A.~Denner, and S.~Dittmaier,
\newblock Phys.Rev. {\bf D77}, 013002 (2008), 0710.4749.

\bibitem{Catani:1996vz}
{S. Catani and M.H. Seymour},
\newblock Nucl. Phys. {\bf B485}, 291 (1997), hep-ph/9605323.

\bibitem{Campanario:2011cs}
F.~Campanario,
\newblock JHEP {\bf 1110}, 070 (2011), 1105.0920.

\bibitem{Cullen:2011ac}
G.~Cullen {\em et~al.},
\newblock Eur.Phys.J. {\bf C72}, 1889 (2012), 1111.2034.

\bibitem{Denner:2006ic}
A.~Denner and S.~Dittmaier,
\newblock Nucl.Phys.Proc.Suppl. {\bf 160}, 22 (2006), hep-ph/0605312.

\bibitem{vanHameren:2010cp}
A.~van Hameren,
\newblock Comput.Phys.Commun. {\bf 182}, 2427 (2011), 1007.4716.

\bibitem{Passarino:1978jh}
G.~Passarino and M.~Veltman,
\newblock Nucl.Phys. {\bf B160}, 151 (1979).

\bibitem{Denner:2005nn}
A.~Denner and S.~Dittmaier,
\newblock Nucl.Phys. {\bf B734}, 62 (2006), hep-ph/0509141.

\bibitem{Campanario:2011ud}
F.~Campanario, C.~Englert, M.~Rauch, and D.~Zeppenfeld,
\newblock Phys.Lett. {\bf B704}, 515 (2011).

\bibitem{Campanario:2013qba}
F.~Campanario, M.~Kerner, L.~D. Ninh, and D.~Zeppenfeld,
\newblock Phys. Rev. Lett. 111, {\bf 052003} (2013), 1305.1623.

\bibitem{Sjodahl:ColorFull}
M.~Sjodahl,
\newblock work in preparation .

\bibitem{Sjodahl:2012nk}
M.~Sjodahl,
\newblock Eur.Phys.J. {\bf C73}, 2310 (2013), 1211.2099.

\bibitem{Lai:2010vv}
H.-L. Lai {\em et~al.},
\newblock Phys.Rev. {\bf D82}, 074024 (2010), 1007.2241.

\bibitem{Pumplin:2002vw}
J.~Pumplin {\em et~al.},
\newblock JHEP {\bf 0207}, 012 (2002), hep-ph/0201195.

\bibitem{Cacciari:2008gp}
M.~Cacciari, G.~P. Salam, and G.~Soyez,
\newblock JHEP {\bf 0804}, 063 (2008), 0802.1189.

\bibitem{Cacciari:2011ma}
M.~Cacciari, G.~P. Salam, and G.~Soyez,
\newblock Eur.Phys.J. {\bf C72}, 1896 (2012), 1111.6097.

\end{thebibliography}
